# Nonextensive thermodynamic relations based on the assumption of temperature duality

Zheng Yahui[a,b], Du Jiulin[a]

a: Department of Physics, School of Science, Tianjin University, Tianjin 300072, China.
b: Department of Physics, School of Science, Qiqihar University, Qiqihar City 161006, China.

**Abstract**

The nonextensive thermodynamic relations are expressed under the assumption of temperature duality, endowing the "physical temperature" and the "Lagrange temperature" in different physical sense. Based on this assumption, two sets of parallel Legendre transform structures are given. One is called "physical" set, and the other called "Lagrange" set. In these two formalisms, the thermodynamic quantities and the thermodynamic relations are both liked through the Tsallis factor. Application of the two sets of the thermodynamic relations to the self-gravitating system shows that the heat capacity defined in the classical thermodynamics has no relevance to the stability of the system. Instead, the new defined heat capacity can determine the stability of the system.

**Keywords:** Thermodynamic relation; Temperature duality; Nonextensive thermodynamics.

PACS: 05.90.+m; 05.20.Gg; 95.30.Tg.

## 1. Introduction

Thermodynamics is originally derived from the phenomenal observations and experiments. This makes the scientists to believe that a perfect thermodynamic theory should be applicable to any realistic physical system regardless of the statistical basis. In other words, when we try to develop a new statistical theory different from the classical Boltzmann-Gibbs statistics, we always hope all the thermodynamic relations keep the same forms as the traditional ones. However, this idea may be reexamined carefully based on what sense the statistical basis is changed.

When our view is transferred from Boltzmann-Gibbs statistics to the quantum statistics, the basic thermodynamic relations are applicable without any modifications. In recent years, it has been found that many complex systems including the systems with long range interactions, fractal space-time and the non-Markov process can not be described by the traditional statistics. These complex systems are indivisible and nonextensive. It has been recognized that the classical statistics can only describe the systems where the short-range interactions are dominant. Therefore, the classical thermodynamic relations should be generalized to the nonextensive systems. In fact, a modified thermodynamic formalism for the nonextensive systems can be constructed on the basis of nonextensive statistical mechanics (NSM), which based on the nonextensive entropy proposed firstly by Tsallis in 1988 [1]. By use of this new statistical theory, a power-law distribution function can be easily deduced. So far, NSM has been applied to many scientific research fields, such as plasma physics [2], astrophysics [3], anomalous diffusion [4], biological physics [5], econophysics [6], and so on. Simultaneously, the thermodynamic relations in NSM have also been developed [7,8].

Even though, a fundamental problem is remained in the effort to develop an appropriate thermodynamic formalism for the nonextensive systems. This is historically related to the problem of the zeroth law of thermodynamics. The core of the zeroth law problem is how to define a state variable like temperature to describe the "equilibrium" state of a nonextensive system. However, the key is that, no matter how to define the so-called physical temperature [8], it is impossible to correspond to the "temperature" existing in a realistic nonextensive system. That is because such a nonextensive system is always at nonequilibrium state, where the "temperature" is inhomogeneous, much different from the homogeneous physical temperature. In order to solve the problem, recently we proposed an assumption of temperature duality [9]. We gave a dual explanation for the concept of temperature in NSM. That is, in the domain of NSM, the concept of temperature may be split into two kinds of explanation. One is the physical temperature defined at the 'equilibrium' state with the standard ensemble method [8]. The other is the Lagrange temperature, defined through the Legendre relation or the Lagrange multiplier [9]. The physical temperature can be homogeneous, but the Lagrange temperature can be inhomogeneous in the nonextensive systems such as self-gravitating systems. These two concepts of temperature are linked through the Tsallis factor. This assumption of temperature duality can well solve the difficulty in the definition about temperature in NSM, so as to make NSM applicable for the systems with nonequilibrium property.

In physics, the physical temperature is related to the global property of a nonextensive system, which exhibits as the long range correlation or interaction in phase space or space time in different systems. The long range nature produces a non-localized effect when one tries to measure the physical temperature. That is, the potential energy induced by the long range nature can not be felt by a thermometer; therefore, when we try to use such a thermometer to measure the physical temperature, it is impossible to get its correct value from the thermometer. In this sense, we judge that the physical temperature is nonmeasurable in experiments. Contrary to this, the Lagrange temperature is given for the local physical process which is related to the short-range interactions between particles (such as the molecular collision) and then it is measurable in experiments.

In this work, based on the assumption of temperature duality, we will use NSM as the basis to generalize the classical thermodynamic formalism to the nonextensive one. This may be useful to understand the thermodynamic quantities and the thermodynaics relations in complex systems. It can be expected that two parallel thermodynamical formalisms would be presented ultimately.

## 2. Expressions of the zeroth law of thermodynamics

Before re-discussing the zeroth law of thermodynamics, let us give some basic definitions and deductions. The central concept of the NSM is Tsallis entropy [1], which is expressed as,

$$S_q = k \frac{\sum p_i^q - 1}{1-q} \equiv k \frac{C_q - 1}{1-q}, \tag{1}$$

where $p_i$ is the probability of the $i$th microscopic state of the system, and the $C_q$ is the Tsallis factor which is very useful in the current work. The quantity $q$ is a real number, called nonextensive parameter, while $k$ is the Boltzmann constant. Tsallis entropy is nonextensive and obeys the following nonadditive rule,

$$S_q(A+B) = S_q(A) + S_q(B) + \frac{1-q}{k} S_q(A) S_q(B). \tag{2}$$

The symbols *A* and *B* denote the subsystems of considered nonextensive system. Another concept we should present for our discussions is the internal energy. Now that we want to establish two parallel thermodynamic formalisms, two parallel internal energies are proposed as follows [10], with *i*th state energy $\varepsilon_i$ and $\varepsilon_i'$ respectively,

$$U_q^{(2)} \equiv \sum p_i^q \varepsilon_i , \tag{3}$$

$$U_q^{(3)} \equiv \frac{\sum p_i^q \varepsilon_i'}{\sum p_i^q} . \tag{4}$$

Obviously the internal energy (3) has ever been defined and then abandoned. In the following discussions, we will endow it with completely different sense. It is notable that the two microscopic state energy levels in (3) and (4) yield to different addition rules, that is,

$$\varepsilon_{i,j}(A,B) = \varepsilon_i(A) + \varepsilon_j(B) - (1-q)\beta'\varepsilon_i(A)\varepsilon_j(B) , \tag{5}$$

$$\varepsilon_{i,j}'(A,B) = \varepsilon_i'(A) + \varepsilon_j'(B) - (1-q)\frac{\beta}{C_q}(\varepsilon_i'(A) - U_{qA}^{(3)})(\varepsilon_j'(B) - U_{qB}^{(3)}) . \tag{6}$$

With the definitions (3) and (4), the distribution functions are given by

$$p_i^{(2)} = \frac{1}{Z_q^{(2)}}[1-(1-q)\beta'\varepsilon_i]^{1/(1-q)}$$
$$Z_q^{(2)} \equiv \sum [1-(1-q)\beta'\varepsilon_i]^{1/(1-q)} \tag{7}$$

$$p_i^{(3)} = \frac{1}{\overline{Z}_q^{(3)}}[1-(1-q)\frac{\beta}{C_q}(\varepsilon_i' - U_q^{(3)})]^{1/(1-q)}$$
$$\overline{Z}_q^{(3)} \equiv \sum [1-(1-q)\frac{\beta}{C_q}(\varepsilon_i' - U_q^{(3)})]^{1/(1-q)} \tag{8}$$

The quantity $\beta$ is the Lagrange multiplier and $\beta'$ is the generalized Lagrange multiplier, whose meaning will be explained in the next context.

Here, a condition is acquiescent, namely, the independence assumption of probability,

$$p_{i,j}(A,B) = p_i(A)p_j(B) . \tag{9}$$

According to the addition rule of microstates (5), it is easy to verify that the distribution function (7) satisfies the above independence assumption. In the addition rule (6), different from the expression given in the references [8,11,12,13], we add a cross term in this rule, which indicates the microstate energy level is also non-additive. However, it is apparent that the macroscopic energy is additive in the light of (4) and (6), i.e.,

$$U_q^{(3)}(A,B) = U_q^{(3)}(A) + U_q^{(3)}(B) . \tag{10}$$

That is because the summation of the probability distribution, in the energy constraint (4), makes the cross term disappear. Then, in view of (6) and (10) one can further verify that

$$[1-(1-q)\frac{\beta}{C_q}(\varepsilon_i' - U_{qA}^{(3)})][1-(1-q)\frac{\beta}{C_q}(\varepsilon_j' - U_{qB}^{(3)})]$$
$$= [1-(1-q)\frac{\beta}{C_q}(\varepsilon_{ij}' - U_{qAB}^{(3)})] \tag{11}$$

Here, we set that

$$\frac{\beta}{C_q}(A,B) = \frac{\beta}{C_q}(A) = \frac{\beta}{C_q}(B), \tag{12}$$

which can be proved in the following context. Actually, the generalized Lagrange multiplier in (5) satisfies the similar relation. It is apparent that, according to (8) the equation (11) leads to the independence assumption (9).

Now we reexamine the zeroth law of thermodynamics in NSM. There are two aspects about the zeroth law problem: the first is how to define a homogeneous state-variable like temperature; the second is how to apply it to nonequilibrium complex systems. Many authors have done enough discussions about the first aspect in different cases. For example, Abe analyzed the Tsallis equilibrium of statistical ensemble, defined the so-called physical temperature and applied it to an ideal gas model [8,11]. Martinez et al tried to reconcile the nonextensive statistics with the zeroth law by recourse to a modified Lagrange multiplier [12]. Johal deduced an additive entropy form by use of algebraic method [13]. Chen and Ou obtained the normal Legendre relation based on the isothermal assumption and the independence assumption of probability [14]. Scarfone explored an addition rule of the energy level of microstate in use of algebraic analysis [15].

However, the second aspect has never been noticed before. In order to settle the second aspect about zeroth law problem, we have presented the assumption of dual temperatures [9]. We gave a dual explanation to the concept of temperature: one is the physical temperature and another is the Lagrange temperature. The physical temperature is the state variable to describe the Tsallis equilibrium state, while the Lagrange temperature could be inhomogeneous at the same state. Apparently our scheme meets these two aspects meanwhile. Yet, there is no doubt that this idea of dual temperature needs to be explored further. In the following we will reexamine the zeroth law of thermodynamics to further expound the dual explanations to the concept of temperature in the domain of NSM.

For an isolated system at 'equilibrium' state (the Tsallis equilibrium), both the differentials of (2) and (10) are zero. Therefore, one has

$$dS_q(A,B) = C_q(B)\frac{\partial S_q(A)}{\partial U_q^{(3)}(A)} dU_q^{(3)}(A) + C_q(A)\frac{\partial S_q(B)}{\partial U_q^{(3)}(B)} dU_q^{(3)}(B) = 0, \tag{13}$$

$$dU_q^{(3)}(A,B) = dU_q^{(3)}(A) + dU_q^{(3)}(B) = 0. \tag{14}$$

The above two equations lead to

$$\frac{1}{C_q(A)}\frac{\partial S_q(A)}{\partial U_q^{(3)}(A)} = \frac{1}{C_q(B)}\frac{\partial S_q(B)}{\partial U_q^{(3)}(B)}. \tag{15}$$

On the other hand, it is easy to obtain from (4) and (8) that [10] (see Appendix),

$$C_q = [\overline{Z}_q^{(3)}]^{1-q}. \tag{16}$$

Then from this result one can deduce

$$\frac{\partial S_q}{\partial U_q^{(3)}} = k\beta \equiv \frac{1}{T}. \tag{17}$$

The quantity $T$ defined in above equation is the so-called Lagrange temperature. The definition in the above equation (17) is called Lagrange relation.

Generally, in most of nonextensive systems the long-range correlation in phase space or space time plays an important role. This means the energy related to long-range correlation or interaction, namely the potential energy, is significant now. Naturally this presents a dual explanation to the concept of internal energy. In the classical ideal gas, the internal energy refers to the sum of the kinetic energies of all molecules in system. In such system, the 'thermal balance' is achieved only through the heat exchange between different parts. Here the 'heat' refers to the transfer of kinetic energy of molecules. So the thermometer can feel the 'temperature' in the classical system: the reading on the thermometer is determined by the heat it gets or loses.

In the nonextensive systems, the meaning of "heat" is remained. However, the thermal balance can only hold in the local sense. That is because only in the local part the effect of long range nature can be ignorable and the potential energy related to long range nature does not participate the energy exchange in the small scale. When we put a thermometer into a nonextensive gaseous system, such as self-gravitating gaseous system, it can only be in contact with a very small volume of molecules. And the thermometer can only exchange "heat" with this small volume, through which a equilibrium between the thermometer and the small volume is achieved ultimately; so what the thermometer measures is actually the absolute temperature; it 'reads out' the value of Lagrange temperature dependent on how much "heat" it gets or loses. Correspondingly, now that the thermometer can not feel the long range potential energy of the small volume, the internal energy of this small volume should be the sum of kinetic energies of molecules inside it (maybe including some intermolecular potential energy). That is, the classical concept of "internal energy" holds in the nonextensive systems.

On the other hand, it is universally accepted that a nonextensive system is indivisible. This indicates that the "internal energy" in such a system is nonadditive. Fundamentally, the equation (17) does not hold in the whole system, now that the "internal energy" (4) is additive, see (10). However, if we focus our attention on the small scale, the "internal energy" (4) could be explained as the sum of kinetic energies of molecules (and a few intermolecular potential energy) in the small scale. Then equation (17) holds in this situation, and the Lagrange temperature is regarded as the absolute temperature in classical statistics [16]. It is measurable in experiments. The Lagrange temperature is essential to settle the second aspect of zeroth law problem; therefore the internal energy (4) is also essential. In order to differ from the "internal energy" in (3), we call the one in (4) Lagrange internal energy.

Considering the Lagrange relation (17), the equation (15) becomes

$$\frac{1}{C_q(A)T(A)} = \frac{1}{C_q(B)T(B)}. \tag{18}$$

We can see that the Lagrange temperature is ordinarily inhomogeneous in the whole system (the Tsallis factor is also inhomogeneous meanwhile). On the other hand, the zeroth law of thermodynamics demands a homogeneous state-variable whose role is like the temperature concept. In order to resolve this problem, the physical temperature $T_q$ is defined as [8,11,16]

$$T_q \equiv C_q T. \tag{19}$$

Immediately, one gets

$$T_q(A) = T_q(B). \tag{20}$$

This means the "physical temperature" is homogeneous at the equilibrium in an isolated system. Obviously this temperature is responsible for the "balance" between different parts of system. In this sense, the definition (19) resolves the first aspect about the zeroth law problem. Here the "balance" is not the "thermal balance", but the "energy balance". That is because in global sense the long range potential energy participates in the energy exchange. Just as mentioned above, when put a thermometer into a nonextensive system, it can not feel the long range potential energy. Therefore, the thermometer is unable to feel the energy balance and the measurement to the "physical temperature" is unrealistic.

Here one difficulty is how to judge whether the energy balance is achieved or not, now that the physical temperature is nonmeasurable. One criterion is whether there is energy dissipation in the system or not. If we can not detect any energy dissipation, such as heat conduct and radiative transfer, the system must be at energy balance. From the viewpoint of molecular kinetic theory, zero energy dissipation indicates the total energy of molecules (averaged kinetic energy, potential energy and radiation energy stored in molecules) is a constant everywhere. This is the dynamic explanation of energy balance [17]. From standpoint of evolution, such a system with energy balance already evolves to its ultimate state. In other words, this state can be regarded as the extremum state of some state function such as entropy. The generalized *H*-theorem [18] guarantees this state is at the minimum of the generalized *H*-function (the minus of Tsallis entropy), in which condition the detailed balance must be satisfied. Ordinarily, we call the ultimate state of evolution Tsallis equilibrium state. The generalized *H*-theorem presents a manner linking the energy balance to detailed balance. The energy balance, together with the detailed balance has become the statistical basis of physical temperature. Furthermore, it should be noticed that from (20) we can easily deduce (12), which ensures in logic the independence assumption of probability (9).

The equation (19) is an expression of the assumption of temperature duality, which can link the physical temperature $T_q$ to the Lagrange temperature $T$ through Tsallis factor (1). Now that physical temperature is un-measurable while the Lagrange temperature is measurable in experiments, the equation (15) seems to show that the un-measurability is associated with the Tsallis factor. On the other hand, equation (1) indicates that Tsallis factor is dependent on the nonextensive parameter. Therefore, it is reasonable to judge that the un-measurability should be related to the nonextensivity.

In the above discussion, we define the physical temperature in an indirect manner. Next, we would present a direct manner to redefine this state variable by aid of (3). Apparently, the internal energy defined in (3) yields to the nonadditive rule, that is [19],

$$U_q^{(2)}(A+B) = C_q(B)U_q^{(2)}(A) + C_q(A)U_q^{(2)}(B) - (1-q)\beta' U_q^{(2)}(A)U_q^{(2)}(B). \qquad (21)$$

The nonadditivity of this type of internal energy is interesting in physics. It reflects the essence of the interaction existing in nonextensive systems. We would find that with the nonadditive energy a state variable like temperature can be defined in more direct manner. Actually, if we assume an appropriate form of nonadditivie energy, we can always find the expected intensive variables such as temperature and pressure [20]. In the representation of (3), the Tsallis factor can be written as

$$C_q = [Z_q^{(2)}]^{1-q} + (1-q)\beta' U_q^{(2)}. \qquad (22)$$

Considering still the equilibrium state of an isolated system, there are

$$dS_q(A+B) = C_q(B)\frac{\partial S_q(A)}{\partial U_q^{(2)}(A)}dU_q^{(2)}(A) + C_q(A)\frac{\partial S_q(B)}{\partial U_q^{(2)}(B)}dU_q^{(2)}(B) = 0, \qquad (23)$$

$$dU_q^{(2)}(A+B) = C_q(B)dU_q^{(2)}(A) + C_q(A)dU_q^{(2)}(B) = 0. \qquad (24)$$

Combining these two equations, we have

$$\frac{\partial S_q(A)}{\partial U_q^{(2)}(A)} = \frac{\partial S_q(B)}{\partial U_q^{(2)}(B)},\qquad(25)$$

One can easily find from (22) that

$$\frac{\partial S_q}{\partial U_q^{(2)}} = k\beta' \equiv \frac{1}{T_q'}.\qquad(26)$$

This definition can be called a generalized Lagrange relation, according to which we can get from equation (25) that

$$T_q'(A) = T_q'(B).\qquad(27)$$

This indicates that the temperature $T_q'$ defined in (26) plays the same role as the "physical temperature" $T_q$ in (19). They have the similar characteristic in physics, i.e., the one in (26) is also homogeneous at the Tsallis equilibrium state. It is also responsible for the global thermal balance of the system. Therefore in physics it is reasonable to let that

$$T_q' \equiv T_q.\qquad(28)$$

This equivalence directly results in

$$\frac{\partial S_q}{\partial U_q^{(2)}} = \frac{1}{T_q}.\qquad(29)$$

This generalized Legendre relation shows the relation between the "physical temperature" and the internal energy defined in (3). On the other hand, the equation (24) indicates the internal energy (3) is nonadditive and this property is derived from the Tsallis factor. This strongly indicates that the internal energy (3) is nonmeasurable in experiments. It may be called the "physical internal energy". Now that the statistical basis of physical temperature is the energy balance [17], the physical internal energy could be recognized as the sum of potential energy, kinetic energies of molecules and some of radiation energy.

In light of (17) and (19), the equivalence (28) means that

$$\beta' = \frac{\beta}{C_q}.\qquad(30)$$

This relation is interesting and useful. Additionally, according to the above equation and comparing (17) and (29), we have that

$$\frac{\partial S_q}{\partial U_q^{(3)}} = C_q \frac{\partial S_q}{\partial U_q^{(2)}}.\qquad(31)$$

In view of the definitions (3) and (4), the following equivalence is proposed,

$$U_q^{(2)} \equiv C_q U_q^{(3)}.\qquad(32)$$

What should be emphasized here is that, in light of equations (5) and (6), the above equation is not the natural mathematical result of (3) and (4) (they have different microstate energy levels). The equivalence (32) shows that the "physical internal energy" and the "Lagrange internal energy" are also linked through the Tsallis factor.

In this section, we tried to tackle the problem of the zeroth law in the domain of NSM. In the previous works [11-15], they defined only one state-variable like temperature through various methods. This is the so-called physical temperature. In recent work [21], Biró and Ván verified that when given additive energy, in order to directly define the physical temperature we must adopt the additive entropy. Actually, if given additive energy and additive entropy, or nonadditive energy and nonadditive entropy, the physical temperature can always be defined directly.

However, there is no counterpart about it in the nonextensive systems, because the realistic temperature in the nonextensive system is always inhomogeneous but the defined temperature on the basis of the zeroth law is homogeneous. So the dual definitions about the concept of temperature are needed. By recourse to the dual explanations, we have then obtained the relation (32), where the Tsallis factor links the "physical internal energy" to the "Lagrange internal energy". It is very similar to the structure of the expression (19) of the assumption of the duality temperature. It is expected that other quantities can also obey the same structure.

## 3. Expressions of the first law of thermodynamics

For getting other thermodynamic relations, let us examine the first law of thermodynamics in the representations of (3) and (4). By use of the method in [22], from (3)-(4) and (7)-(8) we can find that,

$$dU_q^{(3)} = TdS_q - dW = TdS_q - PdV ,\qquad(33)$$

$$dU_q^{(2)} = T_q dS_q - dW' = T_q dS_q - P_q dV .\qquad(34)$$

The quantity $W$ is the work done by the system to the circumstance. In the deduction of (34) the equivalence (28) has been applied. In the above two equations, we give two pressures, $P$ and $P_q$, which have different physical meanings. In light of (31), or in view of the equalities (15) and (24), we obtain,

$$dU_q^{(2)} = C_q dU_q^{(3)} ,\qquad(35)$$

and we can get the relation between the these two pressures,

$$P_q = C_q P .\qquad(36)$$

It is notable that this is also similar to (19). Now that the Lagrange internal energy is measurable, the work related to it should be also measurable; then the pressure in (33) is measurable and it may be called the "Lagrange pressure". Similarly the pressure in (34) is un-measurable and it may be called the "physical pressure".

Now let us make the Legendre transformation. We define two kinds of free energy as follows,

$$F_q^{(3)} = U_q^{(3)} - TS_q ,\qquad(37)$$

$$F_q^{(2)} = U_q^{(2)} - T_q S_q .\qquad(38)$$

Further one can find,

$$F_q^{(2)} = C_q F_q^{(3)} .\qquad(39)$$

This is another similar structure. It is obvious that due to the non-additive Tsallis entropy, these two free energies are both non-additive. However, Tsallis entropy can be calculated through the Lagrange relation (17), after the measurements of the Lagrange temperature and internal energy. Therefore, the free energy in (37) is thought to be a measurable quantity. And the free energy in (38) is un-measurable, now that the "physical internal energy" and the "physical temperature" are un-measurable. Based on such a consideration we may call the free energy in (37) the "Lagrange free energy", and the one in (38) the "physical free energy". In view of (22), one can further get

$$F_q^{(2)} = -kT_q \ln_q Z_q^{(2)} \equiv -kT_q \frac{(Z_q^{(2)})^{1-q} - 1}{1-q} ,\qquad(40)$$

where the relation (30) has been considered. Considering the relation (39), we have

$$F_q^{(3)} = -kT \ln_q Z_q^{(2)} .\qquad(41)$$

Generally, we can define a new partition function in representation (4), which [10,23] is

$$\ln_q \bar{Z}_q^{(3)} = \ln_q Z_q^{(3)} + \beta U_q^{(3)} .\qquad(42)$$

Comparing it with (22) and in light of (16) we have

$$Z_q^{(3)} \equiv Z_q^{(2)} .\qquad(43)$$

So in our discussions, we omit the definition of the new partition function (42).

Considering the total derivatives of (37) and (38), one would obtain

$$dF_q^{(3)} = -PdV - S_q dT, \quad (44)$$

$$dF_q^{(2)} = -P_q dV - S_q dT_q. \quad (45)$$

Obviously, there is,

$$S_q = -\left(\frac{\partial F_q^{(3)}}{\partial T}\right)_V = -\left(\frac{\partial F_q^{(2)}}{\partial T_q}\right)_V = -\left(\frac{\partial (C_q F_q^{(3)})}{\partial (C_q T)}\right)_V. \quad (46)$$

The above equation shows that when the whole volume of a system is fixed, Tsallis factor also does not change. Now that Tsallis factor is related to the nonextensiveity, a reasonable conclusion is that the nonetensivity should be related to the configuration of the system. Furthermore, we get

$$P = -\left(\frac{\partial F_q^{(3)}}{\partial V}\right)_T, \quad (47)$$

$$P_q = -\left(\frac{\partial F_q^{(2)}}{\partial V}\right)_{T_q} = -\left(\frac{\partial (C_q F_q^{(3)})}{\partial V}\right)_{T_q}. \quad (48)$$

The expression (48) again shows that the un-measurability of the "physical pressure" is derived from the Tsallis factor or the nonextensivity. In view of (32), one further gets

$$\left(\frac{\partial (C_q F_q^{(3)})}{\partial V}\right)_{T_q} = C_q \left(\frac{\partial F_q^{(3)}}{\partial V}\right)_T. \quad (49)$$

It is notable that Tsallis factor can not be cancelled out from two sides of this identity.

According to the distribution function (7), it is obvious that

$$U_q^{(2)} = -\frac{\partial}{\partial \beta'} \ln_q Z_q^{(2)}. \quad (50)$$

And then, considering the equivalence (32), there is

$$U_q^{(3)} = -\frac{\partial}{\partial \beta} \ln_q Z_q^{(2)}. \quad (51)$$

The above partial derivative is given for a constant volume. According to (46), Tsallis factor is invariable in this situation; so Tsallis factor can be extracted from the partial derivative in (50). This equation (51) gives a method to calculate the partition function in (7). Through the partition function, we can get some information about this distribution function (7). Now we define the heat capacity with a fixed volume, namely,

$$C_V^{(3)} = \left(\frac{\partial U_q^{(3)}}{\partial T}\right)_V, \quad (52)$$

$$C_V^{(2)} = \left(\frac{\partial U_q^{(2)}}{\partial T_q}\right)_V. \quad (53)$$

Now that they are defined in fixed volume, these two heat capacities are equivalent according to (46). So for simplicity we ordinarily define

$$C_V^{(2)} = C_V^{(3)} \equiv C_V = -\frac{\partial^2 \ln_q Z_q^{(2)}}{\partial T \partial \beta} = -\frac{\partial^2 \ln_q Z_q^{(2)}}{\partial T_q \partial \beta'}. \quad (54)$$

The generalized form of heat capacity is useful in the nonextensive thermodynamics, which is given by

$$C_{gV} = \left(\frac{\partial U_q^{(2)}}{\partial T}\right)_V = C_q \left(\frac{\partial U_q^{(3)}}{\partial T}\right)_V = C_q C_V . \tag{55}$$

About the significance of the above relation, we will give a complete analysis in section 4.

Now we can further define the other thermodynamic quantities, such as the enthalpy,

$$H_q^{(3)} = U_q^{(3)} + PV , \tag{56}$$

$$H_q^{(2)} = U_q^{(2)} + P_q V . \tag{57}$$

In the same way, we give the enthalpy (56) a name "Lagrange enthalpy", which is measurable. And the one in (57) may be called the "physical enthalpy" that is un-measurable. They are also linked through Tsallis factor, that is,

$$H_q^{(2)} = C_q H_q^{(3)} . \tag{58}$$

Furthermore we calculate out that

$$dH_q^{(3)} = TdS_q + VdP , \tag{59}$$

$$dH_q^{(2)} = T_q dS_q + VdP_q . \tag{60}$$

It is easy to obtain the following relations,

$$T = \left(\frac{\partial H_q^{(3)}}{\partial S_q}\right)_P , \quad T_q = \left(\frac{\partial H_q^{(2)}}{\partial S_q}\right)_{P_q} , \tag{61}$$

$$V = \left(\frac{\partial H_q^{(2)}}{\partial P_q}\right)_{S_q} = \left(\frac{\partial H_q^{(3)}}{\partial P}\right)_{S_q} = \left(\frac{\partial (C_q H_q^{(3)})}{\partial (C_q P)}\right)_{S_q} \tag{62}$$

The equation (62) shows that, like the invariance with volume (46), the Tsallis factor also does not change with fixed Tsallis entropy. Likewise, the "Lagrange Gibbs-function" and the "physical Gibbs-function" can also be fined as

$$G_q^{(3)} = U_q^{(3)} - TS_q + PV , \tag{63}$$

$$G_q^{(2)} = U_q^{(2)} - T_q S_q + P_q V . \tag{64}$$

And then we find

$$G_q^{(2)} = C_q G_q^{(3)} , \tag{65}$$

$$dG_q^{(3)} = -S_q dT + VdP , \tag{66}$$

$$dG_q^{(2)} = -S_q dT_q + VdP_q . \tag{67}$$

Up to now, we can claim that two sets of parallel thermodynamic formalisms in NSM are developed. The "Lagrange set" is measurable, and the "physical set" is un-measurable. They are linked through Tsallis factor in (1). In each set, we can obtain the completely similar nonextensive thermodynamic relations as those in the classical thermodynamics, including the famous Maxwellian relations. Here we obtain two parallel formalisms of Legendre transformation in the NSM framework.

## 4. The applications to the nonextensive systems

As an application, firstly let us see what we can get when the duality idea is used to an ideal gas. When NSM theory is applicable in the ideal gaseous system, the system should exhibit some nonextensivity. So it is not an 'ideal' gas in the classical statistics again. We may call it a nonextensive gas [24]. Before the discussion, a question should be clarified: what is the physical origin of the nonextensivity in a nonextensive gas? This gas model is apparently a generalization of the classical ideal gas model. A reasonable explanation is that the nonextensivity is originated from the interactions between the molecules which can not be ignored in experiments. These interactions are related to the Joule coefficient, second virial coefficient, and so on. Therefore, the nonextensivity may be regarded as a result from the interactions between molecules, which ordinarily are the short-range molecular forces. Based on such consideration, we can apply the nonextensive gas model to the realistic gas system.

Here, we consider the gas system at ordinary (Lagrange) temperature and pressure. Under this situation, interactions between molecules are not so obvious; therefore the nonextensivity is weak enough. This means $q \approx 1$. In order to calculate the internal energy in the nonextesnive gas, we should know the "physical partition function" in (7), according to (50) and (51). It is difficult to directly calculate this partition function. However, due to $q \approx 1$, the "physical partition function" in (7) has the structure similar to the "Lagrange partition function" in (8), which is given in [8] by

$$\overline{Z}_q^{(3)} = \frac{\Gamma(\frac{2-q}{1-q})}{\Gamma(\frac{2-q}{1-q}+\frac{DN}{2})} \frac{V^N}{N!h^{DN}} [\frac{2\pi m C_q}{(1-q)\beta}]^{DN/2} [1+(1-q)\frac{U_q^{(3)}\beta}{C_q}]^{1/(1-q)+DN/2}, \qquad (68)$$

where $N$ is the molecule number in the gas, $h$ is the linear dimension of elementary cell in phase space, $m$ is the molecular mass, $D$ is the molecular freedom degree, and $\Gamma(x)$ is the gamma function. Of course, this expression is based on the special choice of the distribution function. Here, The reader can see Ref.[8] for detail. Comparing the definitions of these two partition functions in (7) and (8), and considering the equivalence (30), one easily finds that

$$Z_q^{(2)} \simeq \overline{Z}_q^{(3)}\Big|_{U_q^{(3)}=0} = \frac{\Gamma(\frac{2-q}{1-q})}{\Gamma(\frac{2-q}{1-q}+\frac{DN}{2})} \frac{V^N}{N!h^{DN}} [\frac{2\pi m}{(1-q)\beta'}]^{DN/2}, \qquad (69)$$

Here, we ignore the difference between these two energy levels (5) and (6). Noticing that, the above partition function is a function of the volume and the generalized Lagrange multiplier. And according to (50), it is easy to obtain

$$U_q^{(2)} = \frac{D}{2} N k T_q (Z_q^{(2)})^{1-q} \quad . \qquad (70)$$

Now we need the measurable "Lagrange internal energy"; in light of (32), one has

$$U_q^{(3)} = \frac{D}{2} N k T (Z_q^{(2)})^{1-q} \quad . \qquad (71)$$

Considering the relation (65), the above expression is the same as that in [8], also as the result with a kinetic approach [24]. Similarly, according to (40) and (48), we can get,

$$P_q = \frac{NkT_q(Z_q^{(2)})^{1-q}}{V} \quad . \qquad (72)$$

Again, we need the "Lagrange pressure"

$$P = \frac{NkT(Z_q^{(2)})^{1-q}}{V} \quad . \qquad (73)$$

With the expression of partition function (69) (here, due to $q \approx 1$, we let the Tsallis factor in the expression be unity), and by use of the internal energy (71) and pressure (73), one can easily calculate Joule coefficient, second virial coefficient and so on [9,25], recognizing the real gas as the nonextensive gas. The second virial coefficient we calculated in previous work [25] is fitted well with the experimental curve. Lastly, we give an expression of the heat capacity (54),

$$C_V = \frac{D}{2}Nk[1+\frac{D}{2}(1-q)N](Z_q^{(2)})^{1-q} \qquad (74)$$
$$\approx \frac{D}{2}Nk[1+\frac{D}{2}(1-q)N].$$

We consider again the approximation $q \approx 1$ in the above calculation. This heat capacity is different from the result in [24]; it is composed of two apparent parts: one is the heat capacity obtained in the classical ideal gas model, and the other is an additional quantity, relevant to the square of particle number and produced by the nonextensivity of the gas.

Next, let us apply our nonextensive thermodynamic relations into a self-gravitating system. Commonly, in different nonextensive systems the nonextensivity has different physical origins. In a self-gravitating system, the short-range interaction between molecules is ignored and the long-range interaction between them becomes important. So in such a system, the nonextensivity is from the long-range interactions.

In a recent paper [26], we gave an expression of the temperature duality in the molecular dynamics of a self-gravitating system, that is,

$$T_g = T[1+\frac{(q'-1)m\varphi}{kT}] \equiv C_{q'}T, \qquad (75)$$

where the quantity $\varphi$ is the gravitational potential in the mean field, and the quantity $T_g$, called the "gravitational temperature" [26,17], is identical to the "physical temperature" in the ensemble theory. Notice that there is a prime on the nonextensive parameter. This is because the parameter has different meaning from that one in nonextensive gas model. In light of the "gravitational temperature", the corresponding "gravitational heat capacity" [26] is defined as

$$C_V^{(g)} = \frac{dE}{dT_g}. \qquad (76)$$

The quantity $E$ is the total energy as sum of the kinetic energy of molecules and the gravitational interaction potential energy; therefore, it is identical to the "physical internal energy" defined by (3) in the nonextensive thermodynamics. So this heat capacity is recognized as the heat capacity defined in (53), which is identical to the form in the definition (52). Now we need to define the "Lagrange internal energy" which is chosen to be sum of the kinetic energies of the molecules, whose expression is given by

$$U = \frac{D}{2}Nk\bar{T}. \qquad (77)$$

The quantity $\bar{T}$ is the average value of the "Lagrange temperature" for the whole system (the "Lagrange temperature" in self-gravitating system is ordinarily inhomogeneous).

Formally, this internal energy should have the same expression as that one in (71). The reason why we adopt this form (77) is that, the short-range interaction forces between the molecules can be ignored as compared with the long-range gravitational interactions between them; therefore, the nonextensivity originated from short-range interaction disappears in the self-gravitating system. Obviously, according to the viral theorem [27], the total energy, or the "physical internal energy" of the self-gravitating system is

$$E = -U = -\frac{D}{2}Nk\bar{T}. \qquad (78)$$

This is consistent with the result in classical statistics. Then the heat capacity is given by

$$\bar{C}_V = \frac{dE}{d\bar{T}} = -\frac{D}{2}Nk. \qquad (79)$$

Apparently, in the representation of NSM this is the generalized heat capacity defined in (55). Notice that this capacity is negative, which is as a main proof of the thermodynamic instability of a self-gravitating system in the classical statistics [28]. This is not the whole story in fact. Now that the "gravitational temperature" is related to Tsallis equilibrium state, the "gravitational heat capacity" (76) determines the stability of a self-gravitating system [26]. The generalized heat capacity (79), which is 'normal' in classical statistics, is not important to the thermodynamic evolution of a self-gravitating system.

In order to obtain the expression of the "gravitational heat capacity", now we calculate the average of the "gravitational temperature" (75). According to the virial theorem [27], there is,

$$T_g = (1-DQ)\overline{T} \equiv [1-D(q'-1)]\overline{T}. \tag{80}$$

Then according to the two definitions (76) and (79), we have

$$\overline{C}_V = (1-DQ)C_V^{(g)} \equiv \overline{C}_{q'}C_V^{(g)}. \tag{81}$$

This is similar to (55). In view of (79), the expression of the "gravitational heat capacity" is

$$C_V^{(g)} = \frac{\overline{C}_V}{\overline{C}_{q'}} = \frac{D}{2}\frac{Nk}{D(q'-1)-1}. \tag{82}$$

We have proved that [17,29] in a self-gravitating gaseous system, the thermodynamic instability leads to the increase of the nonextensive parameter $q'-1$. This means that the self-gravitating system, although exhibiting complicated evolution characteristics, can approach a stable "Tsallis equilibrium state".

## 5. Conclusions and discussions

The assumption of temperature duality [9] is that the concept of temperature in nonextensive statistics can be split into two parts: the "physical temperature" and the "Lagrange temperature". In this work we have presented two types of the thermodynamic formalisms based on the assumption of temperature duality. One is called physical set, centered on the "physical temperature". The thermodynamic quantities in this set are related to the global property of the system. The other is called Lagrange set, centered on the "Lagrange temperature", whose thermodynamic quantities are related to the local property of the system. These two sets are linked through the Tsallis factor.

Through the Legendre transformations, we have separately defined the freedom, the enthalpy and the Gibbs function under the two sets of formalisms. According to our method, the partition function in the physical set is identical to the newly defined one in the Lagrange set, see Eq.(43). This makes the calculation of the several thermodynamic quantities easy since this partition function is easy to obtain through (7). The equivalences (28) and (32), together with the assumption of temperature duality, are the theoretical basics in the construction of the two sets of formalisms.

The main aim to construct the thermodynamic quantities in the parallel sets of formalisms is that, we try to present the applicable thermodynamic relations in the practical nonextensive complex systems. Among these two sets of formalisms, the physical set is derived due to the inherent theoretical requirement. The Lagrange set of formalism is due to the experimental requirement, with which the nonextensive statistical theory may become applicable in practice. In experiments, we can only obtain the values of the thermodynamic quantities in the Lagrange set, which multiplied by the Tsallis factor become those in the physical set.

Finally, let us give some comments on the definition paradigm of the so-called physical internal energy in Eq.(3). From the definition, a peculiar property can be drawn, that is $\sum 1 p_i^q \neq 1$. This property looks like unreasonable. However, just as pointed by Tsallis [23], this property "*is not easy to interpret, unless one is ready to accept that the entropy nonextensivity results in an effective gain or loss of norm*". This strongly suggests that the nonextensivity affects the measurement to any thermodynamic quantities. It leads to the gain or loss of, say, the internal energy, when one tries to measure a part of a system considered. This is reasonable, since some energy in fact exists in the interface between the measurable part and the other part of the system that can not appear in the measurement.

## Acknowledgements

This work is supported by the National Natural Science Foundation of China under Grant No.11175128, by the Heilongjiang Province Education Department Science and Technology Research Project under No.12541883 and also by National Natural Science Foundation of China under Grant No.11405092.

## Appendix A

The deduction of Eq. (16):

For our aim, we need (3) and (8), which are given by

$$U_q^{(3)} \equiv \frac{\sum p_i^q \varepsilon_i'}{\sum p_i^q} \tag{A.1}$$

$$p_i^{(3)} = \frac{1}{\overline{Z}_q^{(3)}}[1-(1-q)\frac{\beta}{C_q}(\varepsilon_i' - U_q^{(3)})]^{1/(1-q)} \tag{A.2}$$

$$\overline{Z}_q^{(3)} \equiv \sum [1-(1-q)\frac{\beta}{C_q}(\varepsilon_i' - U_q^{(3)})]^{1/(1-q)} \tag{A.3}$$

The Tsallis factor is defined as

$$C_q \equiv \sum p_i^q \tag{A.4}$$

In order to prove this equation (16), we need to prove

$$\sum [1-(1-q)\frac{\beta}{C_q}(\varepsilon_i' - U_q^{(3)})]^{1/(1-q)} = \sum [1-(1-q)\frac{\beta}{C_q}(\varepsilon_i' - U_q^{(3)})]^{q/(1-q)} \tag{A.5}$$

Notice that

$$\sum [1-(1-q)\frac{\beta}{C_q}(\varepsilon_i' - U_q^{(3)})]^{1/(1-q)}$$

$$= \sum [1-(1-q)\frac{\beta}{C_q}(\varepsilon_i' - U_q^{(3)})]^{q/(1-q)} [1-(1-q)\frac{\beta}{C_q}(\varepsilon_i' - U_q^{(3)})] \tag{A.6}$$

$$= \sum [1-(1-q)\frac{\beta}{C_q}(\varepsilon_i' - U_q^{(3)})]^{q/(1-q)}$$

$$-(1-q)\frac{\beta}{C_q}\sum [1-(1-q)\frac{\beta}{C_q}(\varepsilon_i' - U_q^{(3)})]^{q/(1-q)}(\varepsilon_i' - U_q^{(3)})$$

For the second term on the right hand side, there is

$$\sum [1-(1-q)\frac{\beta}{C_q}(\varepsilon_i' - U_q^{(3)})]^{q/(1-q)}(\varepsilon_i' - U_q^{(3)}) = [\overline{Z}_q^{(3)}]^q \sum p_i^q (\varepsilon_i' - U_q^{(3)})$$

$$= [\overline{Z}_q^{(3)}]^q C_q \frac{\sum p_i^q (\varepsilon_i' - U_q^{(3)})}{\sum p_i^q} \tag{A.7}$$

According to (A.1), we have

$$\frac{\sum p_i^q \varepsilon_i'}{\sum p_i^q} - U_q^{(3)} = \frac{\sum p_i^q (\varepsilon_i' - U_q^{(3)})}{\sum p_i^q} = 0 \tag{A.8}$$

So one obtains

$$\sum [1 - (1-q)\frac{\beta}{C_q}(\varepsilon_i' - U_q^{(3)})]^{q/(1-q)} (\varepsilon_i' - U_q^{(3)}) = 0 \tag{A.9}$$

So (A.5) is verified, and then (16) is proved.